# A Multi-stage Collaborative 3D GIS to Support Public Participation

Yingjie Hu [a, b], Zhenhua Lv [a], Jianping Wu [a], Krzysztof Janowicz [b], Xizhi Zhao [a] and Bailang Yu [a]*

[a] *Key Laboratory of Geographic Information Science, Ministry of Education, East China Normal University, Shanghai, P. R. China*

[b] *Department of Geography, University of California Santa Barbara, Santa Barbara, U.S.A.*

This paper presents a collaborative 3D GIS to support public participation. Realizing that public-involved decision making is often a multi-stage process, the proposed system is designed to provide coherent support for collaborations in the different stages. We differentiate *ubiquitous participation* and *intensive participation*, and identify their suitable application stages. The proposed system, then, supports both of the two types of participation by providing synchronous and asynchronous collaboration functionalities. Applying the concept of Digital Earth, the proposed system also features a virtual globe-based user interface. Such an interface integrates a variety of data, functions and services into a unified virtual environment which is delivered to both experts and public participants through the Internet. The system has been designed as a general software framework, and can be tailored for specific projects. In this study, we demonstrate it using a scene modeling case and provide a preliminary evaluation towards its usability.

**Keywords:** public participation; multi-stage decision making; 3D visualization; virtual globe; online collaboration; Digital Earth

*Email: blyu@geo.ecnu.edu.cn

## 1. Introduction

The importance of public participation has been recognized in many decision-making processes, such as urban planning, pollution assessment, and natural resource management (Bailey *et al.* 1999, Innes 2000, Nyerges *et al.* 2006, Dietz and Stern 2008). The involvement of the general public (including stakeholders and interested citizens) can help produce decisions that could better serve people's needs (Al-Kodmany 1999, Joerin *et al.* 2009). Public participation has also been regarded as an approach towards enhancing democratic governance (Rydin and Pennington 2000). In many countries, local authorities are increasingly being required to make planning information publicly accessible to encourage public participation (Hetherington *et al.* 2007, Gordon *et al.* 2011).

A vital goal of public participation is to elicit the local knowledge that could be combined with professional expertise to generate good decisions (Al-Kodmany 1999, Barton *et al.* 2005). Achieving this goal needs public participants to have a good understanding of the project, since an uninformed individual could also input futile information (Gordon *et al.* 2011). Empirical studies show that visualization technologies, such as 2D maps and 3D virtual environments, can facilitate participants' learning and understanding in decision-making, especially spatial decision-making, processes (Al-Kodmany 1999, Simpson 2001, Alshuwaikhat and Nkwenti 2002, Hu *et al.* 2010a). In the past years, 3D visualization has been increasingly used in many public participation studies (Gong and Lin 2006, Howard and Gaborit 2007, Lloret *et*

*al.* 2008, Wu *et al.* 2010, Ki 2011). Established on 3D visualization and Web technologies, virtual globes (e.g., Google Earth) can deliver a fast and seamless virtual environment through the Internet (Craglia *et al.* 2012, Zhang *et al.* 2012), and therefore provides an intuitive and more accessible media for public participation. Meanwhile, virtual globes integrate (geographic) data, models, and web services into a single virtual environment, and offer a unified interface for public participants to explore decision projects.

Public participation also needs to fit the decision-making processes (Thomas 1995, Carson 2009). Most geospatial projects cannot be completed in a single step, and multi-stage analysis and collaborations are often necessary (Dragicevic and Balram 2004, Simao *et al.* 2009). While the exact decision-making processes may vary in different projects, five common stages are generally involved: problem definition, problem analysis, alternative solution generation, alternative solution evaluation, and implementation (Figure 1). These five stages are seldom in a linear sequence, and iterations are often necessary to adjust solutions and overcome potential defects. Ideally, all stakeholders and interested citizens should be able to participate in the first four stages (if not all five stages), since the last stage is more about "implementing" rather than "making" the decision. However, when it comes to practice, public participation often takes different forms that could undermine its effectiveness.

In traditional public participations (e.g., public hearing), spatiotemporal constraints are often presented as a major factor that affects the value of public contributions. Distance factors as well as scheduling conflicts can prevent some interested groups from joining in public meetings (Dragicevic and Balram 2004). Consequently, the collected public opinions may not be as objective as expected. To encourage wider public inputs, some governments publish electronic documents onto the Internet to inform local communities about public issues, bringing the *E-government* (Allen *et al.* 2001). While this form of public participation is primarily one directional (i.e., from the government to the general public), later approaches put more emphasis on bidirectional Web participation and support dynamic interactions between the public and decision makers (Kingston 2002, Rinner *et al.* 2008, Bugs *et al.* 2010).

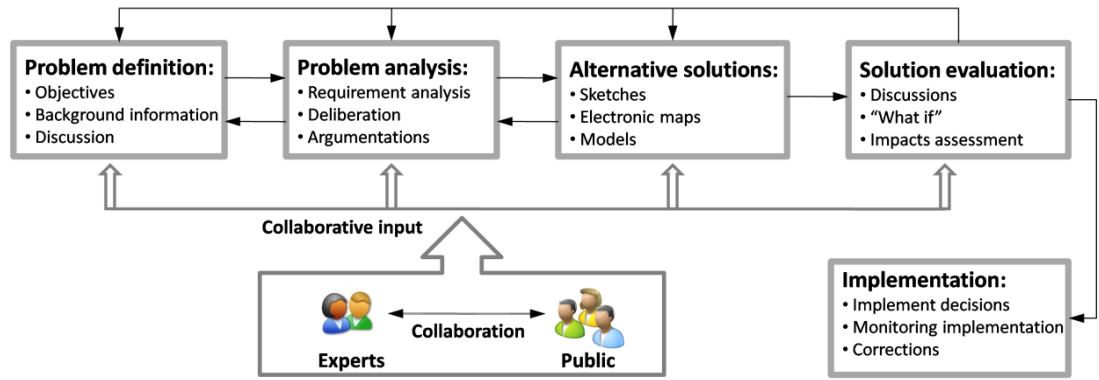

Figure 1: Five stages of a general decision-making process and an idealistic view of public participation

Geographic information systems (GIS) have been widely applied to public participation processes, and *public participation GIS* (PPGIS) was developed (Harris and Weiner 1996, Nyerges *et al.* 1997). The early use of PPGIS was to investigate the participatory processes in face-to-face meetings (Obermeyer 1998). Online PPGIS were developed in following years to overcome the spatiotemporal constraints of traditional public participations (Carver *et al.* 2001). By employing Internet-based asynchronous communication, online PPGIS provides a "24/7" access (i.e., the public have access to the system 24 hours a day and 7 days a week). However, the asynchronous communication also eliminates the real-time back-and-forth dialogs between the public and decision makers, which often produce meaningful results (Gordon *et al.* 2011). From the perspective of multi-stage decision-making, many online participation GIS only enable the public to join in the solution evaluation stage (e.g., reviewing existing solutions and making comments) instead of allowing them to participate in the other decision-making stages (Kingston 2002, Hetherington *et al.* 2007, Wu *et al.* 2010). Synchronous collaborative GIS support real-time interactions among participants, and are often integrated with asynchronous communication components as well as 3D visualization technologies to achieve a more effective public participation (Chang and Li 2008, Klimke and Dollner 2010, Butt and Li 2012).

Employing such an integrated approach, this research presents a collaborative 3D GIS to support public participation in multi-stage decision making. As an application of the concept of Digital Earth, the proposed framework features a virtual globe-based user interface which renders 3D models, map layers, remote sensing images, digital elevation models (DEM), as well as other data in a unified virtual environment. Based on the information-rich 3D environment, both synchronous and asynchronous collaboration modules have been developed to facilitate public participation in different decision-making stages.

The contributions of this paper are summarized as follows:

- Five common stages in general decision making have been identified, and a software framework has been designed to support public participation in these stages.

- We differentiate *ubiquitous participation*, in which participation can take place at any time and any location, and *intensive participation*, which involves back-and-forth conversations and real-time collaborations. Both of them are supported in the proposed software framework.

- A virtual globe has been employed as the primary user interface to facilitate public understanding of the projects, ease the use of the system, and also make the participation process fun.

- The proposed system has been designed as a general framework which can be customized to fit the needs of different projects. A prototype has been implemented as a proof-of-concept, and the source code is shared on GitHub at https://github.com/YingjieHu/PPGIS.

The remainder of this paper is structured as follows. In section 2, we review related works, and identify several development trends in public participation systems. Section 3 discusses the objectives and design considerations. Section 4 presents the system's functionalities and the underlying architecture and key components. A case study is discussed in section 5 to provide a demonstration as well as a preliminary evaluation of its usability. Finally, we summarize the content of this paper and discuss future work.

## 2. Related Works

A quick search in literature reveals related studies not only in PPGIS, but also in collaborative GIS (Churcher and Churcher 1996, Li and Coleman 2003), geocollaboration (Maceachren and Brewer 2004, Cai *et al.* 2005), as well as group-based spatial decision support systems (GSDSS) (Armstrong and Densham 1995, Nyerges 1999, Jankowski and Nyerges 2001). This result is understandable since public participation is also a collaborative and group-based decision-making process. There is also a considerable number of related publications in planning support systems (PSS), as spatial planning is one of the major application areas of public participation (Dragicevic and Balram 2004, Hanzl 2007, Howard and Gaborit 2007, Poplin 2012). While existing systems have been designed for different purposes, we identify several development trends from a technological perspective.

### *2.1 3D Visualization and Virtual Globes*

3D visualization and virtual globes have been increasingly applied to public participation. These technologies ease the understanding of projects for participants (especially non-professional participants), and lower the participation entry (Kreuseler 2000, McCarthy and Graniero 2006, Howard and Gaborit 2007, Hu *et al.* 2010b). Gong and Lin (2006) reported an Internet-based 3D geographic environment and applied it to the collaborative planning of silt dam systems. Employing a virtual globe platform (GeoGlobe), Wu et al. (2010) designed a planning support system which allows public participants to explore urban plans and make their comments. Yiakoumettis et al. (2010) proposed a virtual globe-based framework which enables participants to explore the virtual environment and collaboratively construct 3D models for the planning city region.

*2.2 Geo-referenced Communication and Sketches*

Geo-referenced communication and sketches have been recognized as effective methods to support collaborative processes (Goodchild 2012). Geo-referenced communication can reduce the ambiguity existing in the text comments from participants (Hopfer and Maceachren 2007, Hardisty 2009). The Argumentation Map (Rinner 1998, Rinner 2001, Keßler *et al.* 2005, Rinner *et al.* 2008, Rinner and Bird 2009) shows a good approach to implement geo-referenced communication by explicitly linking a discussion forum and a map display. Sketches have also been suggested as good tools to assist collaborative discussions (Steinmann *et al.* 2005, Tang *et al.* 2005, Warren-Kretzschmar and Tiedtke 2005, Zhao and Coleman 2006). By drawing 2D or 3D geometries (e.g., arrows or cubes), one can visually explain the content in the discussions.

*2.3 An Integration of Synchronous and Asynchronous Collaborations*

While synchronous and asynchronous collaborations have been applied to public participation respectively, there is also a trend in integrating the two. For example, Klimke and Dollner (2010) presented a general model to support both synchronous and asynchronous communication of geo-referenced information in virtual environments. Butt and Li (2012) integrated Web-based GIS and groupware tools into a virtual meeting space (called GeoVPMS) to provide synchronous and asynchronous support of public participation. Computer-supported cooperative work (CSCW) principles have been employed in these systems to support the two types of collaborations, and new modules are being designed to augment the systems' capabilities.

*2.4 Innovations of the Proposed Framework*

The technological development trends and existing public participation systems have shed insights on the proposed multi-stage collaborative 3D GIS. Compared with the related works discussed above, our major innovations can be seen from three perspectives. From the perspective of multi-stage decision making, the proposed framework is designed to cover the different stages of public participation in a more coherent manner. From the perspective of public participation approaches, we identify and combine both *intensive* and *ubiquitous* participations to help elicit meaningful and more objective local knowledge. Finally, from the perspective of technological improvement, we integrates multiple technologies, including virtual globe-based 3D visualization, CSCW, and geo-referenced communication, and present a unified and user-friendly interface to the participants. In the next section, we will discuss the objectives and considerations for the proposed software framework in detail.

**3. Objectives and Design Considerations**

*3.1 Objectives*

The objectives of the collaborative 3D GIS can be summarized as follows:

(1) Providing continuous support to a multi-stage and iterative decision-making process.

(2) Supporting and facilitating the collaborations for geographically distributed participants.

(3) Supporting both intensive and ubiquitous public participation.

(4) Using effective visualizations to ease the understanding of decision projects and make the participation process fun.

*3.2 Design Considerations*

Realizing the above four objectives requires meticulous design considerations. Objective 1 and 3 are related, since different decision-making stages often have different requirements on collaborations, and therefore may prefer different types of participation. As shown in Figure 1, the first three stages need participants to actively discuss the problem and collaboratively work out alternative solutions, which often involve intensive interactions. Therefore, a synchronous collaboration, which supports back-and-forth dialogs and real-time interactions, could help realize intensive participation in these stages. In the solution evaluation stage, the pros and cons of the alternative solutions need to be assessed by a broader range of groups and individuals. For participants who can join in online evaluation sessions, synchronous collaboration allows them to have discussions and get questions answered in a more timely way; for those who cannot be present in the online sessions, asynchronous collaboration offers them a chance to voice their opinions. Therefore, both synchronous and asynchronous collaborations could be provided in this stage to facilitate meaningful and ubiquitous participation.

To achieve objective 2, the collaborative 3D GIS needs to be designed as an Internet-based environment. Such an environment requires a design choice among centralized, replicated, or semi-replicated architecture to handle the distributed data and functions. A centralized architecture uses a central server (or a cluster ) to host the application and data, and controls the input and output of the distributed clients (Brusilovsky *et al.* 1997). A centralized architecture achieves good consistency of shared data, but it also requires a high bandwidth to frequently distribute data to all end users (Chang and Li 2008). A replicated architecture maintains a copy of application and data on each client machine, and synchronize these clients using some mechanisms (Berlage and Genau 1993). Since most functions are executed locally, a replicated architecture reduces bandwidth burden, but also increases the complexity to synchronize multiple copies of shared data. A semi-replicated architecture can balance the pros and cons of the previous two types of architectures (Greenberg and Roseman 1999). This architecture often decomposes an application into several components: some are designed to be shared through the central server, while others are replicated to clients. In our work, a semi-replicated architecture is more suitable since 3D visualization and some other collaboration functions (e.g., sketching tools) need to be executed on the client, while the geospatial and project data need to be centrally maintained to ensure consistency.

For objective 4, virtual globes have been considered as a suitable choice. In addition to the advantages we have discussed in previous sections, some virtual globes, such as Google Earth and Skyline Globe, have provided application programming interfaces (APIs), which can facilitate the integration into customized applications. Recent study also shows that virtual globes can make the participation process fun (Poplin 2012), and therefore could encourage public participation.

Besides the aforementioned four objectives, the proposed system should also establish a concurrency control mechanism since multiple participants would be working together in the same virtual environment simultaneously. Existing methods, such as optimistic concurrency control (OCC), two-phase locking (2PL), user hierarchies, and fixed time quotas, have been applied in database transaction processing (Bernstein *et al.* 1987). In a multi-participant and synchronous collaboration, one person's actions could easily affect the others' (e.g., editing on the same object), which would significantly undermine the performance of OCC. Neither could we adopt a fixed-time-quota strategy for each participant, since the amount of time that a participant may need to complete an action is arbitrary. However, unlike database transaction commitments in which users usually are not aware of the actions of others, a collaborative discussion session allows participants to send messages and exchange ideas. Realizing this difference, we think a leader-and-follower mode, similar to the team-leader mechanism (Maceachren and Brewer 2004), would be a suitable choice. This mode assigns the role of "leader" to one participant and the role of "follower" to all other participants. Only the leader has the privilege to make operations, while followers can send messages to discuss or request for the leader's role. The leader can be an expert, a decision maker or a public participant, and the leader's role can be switched from person to person based on requests. Compared with 2PL and user hierarchies, this mode makes use of the fact that participants can coordinate the collaboration process, thereby avoiding some technical complexities (e.g., deadlock handling).

## 4. Software Architecture and Key Components

The proposed collaborative 3D GIS has been designed as a dual-subsystem architecture (Figure 2) to fit the design considerations. The first subsystem, called "design system", supports the stages of problem definition, problem analysis, alternative solution generation, and a preliminary solution evaluation (the participants are limited to those who can join in the online meetings). The second subsystem, called "review system", allows interested individuals and groups to review the alternative solutions, and opens the evaluation process to a broader public participants.

As shown in Figure 2, this dual-subsystem architecture supports an iterative decision-making process. In the design system, participants can learn background information, discuss project requirements, analyze problems, design alternative solutions, do simulations, and perform evaluations. These processes could be completed in a series of online meetings according to the project requirements, and participants can always return back to a previous process if the current approach does not work out. The alternative solutions are then published into the review system, and are evaluated by a larger number of public participants. The review system stores people's comments in a database which can also be accessed by the design system. After the review process, one or more online meetings can be held using the design system to address people's comments and make revisions. This whole process may take several iterations until an acceptable solution is produced.

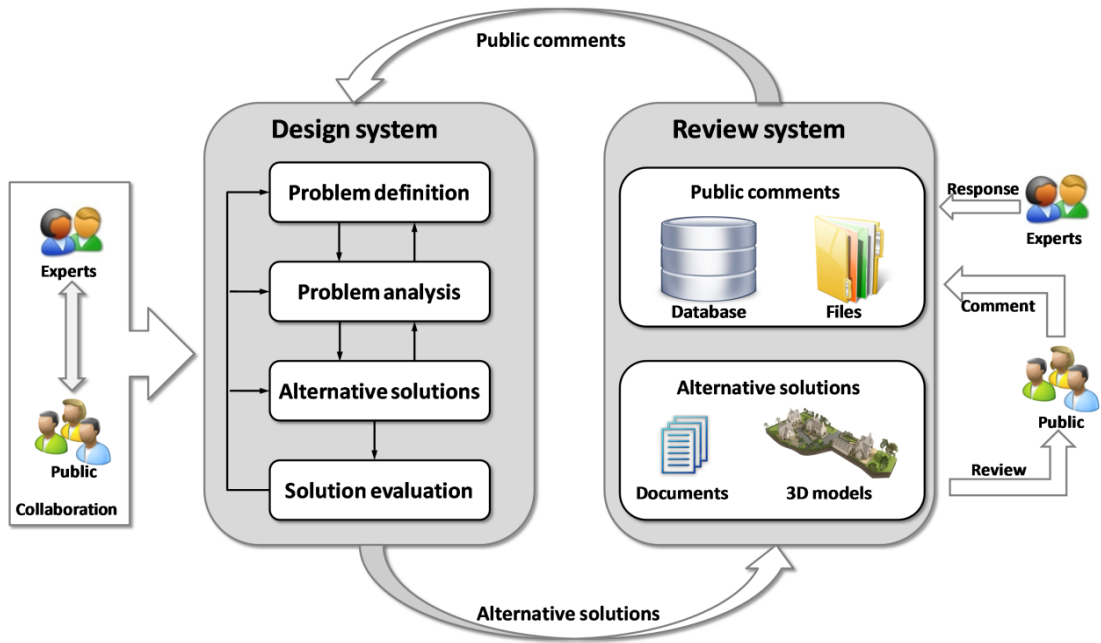

Figure 2: The dual-subsystem architecture and the iteration process

*4.1 The Design System*

The major goal of the design system is to support synchronous collaborations from problem definition to a preliminary evaluation. In this section, we first describe the major functionalities provided by this subsystem, and then present the underlying architectures.

*4.1.1 Functionalities*

There are three major functions in the design system: 3D view sharing, geo-referenced communication, and user operation sharing.

**a) 3D view sharing**

View sharing is a fundamental function for synchronous collaboration. A common view can convey the context under which the discussions, operations and other collaborative activities are taking place (Hardisty 2009, Janowicz 2010, Klimke and Dollner 2010). The function of 3D view sharing has been designed to facilitate the communication of geographically distributed participants. When the leading participant navigates in the virtual environment, all other participants can share his/her view, realizing the effect of "what I see is what you see".

**b) Geo-referenced communication**

While 3D view sharing ensures the same geographic context, ambiguities may still exist in verbal communications. For example, by saying "the tree next to the building", one user is referring to the tree on the left side of the building, while some other participants may interpret it as the tree on the right side. Geo-referenced communication solves this problem by complementing instant messages with sketching tools. It allows participants to draw lines, polygons, arrows, and even text annotations in the virtual environment to identify the corresponding geographic features referred in the instant messages.

**c) User operation sharing**

User operation sharing is designed to help participants learn others' actions, such as performing a spatial analysis or editing an object. This function is important since in an online working environment, participants may not be able to see the actions of others. Therefore, understanding what others are working on could help participants form a clear idea about the current situation. User operation sharing has been integrated to all analysis and collaborative tools to ensure the results are shared to all participants.

*4.1.2 Architecture of the Design System*

To realize the three major functions, a semi-replicated and server-client architecture has been designed (Figure 3).

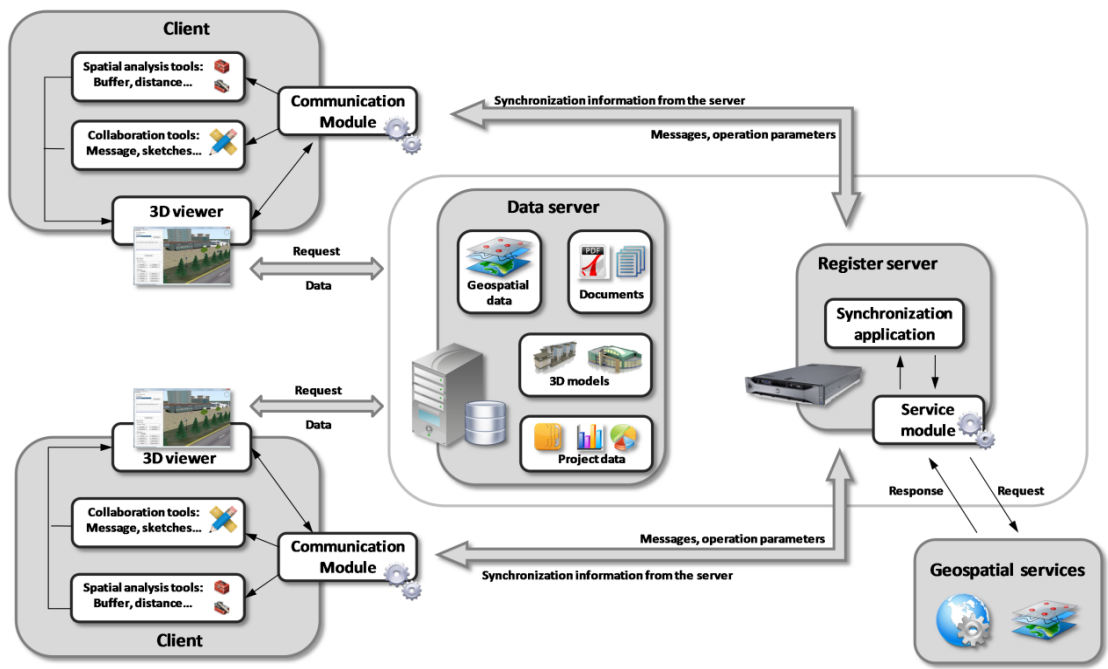

Figure 3: The architecture of the design system (an example of two clients)

The server side contains a data server and a register server. The data server hosts geospatial data, 3D models, documents, and other project data. The register server runs a synchronization application which manages the client connections. A service module is contained in the register server to integrate geospatial tools through external services. This module enables the integration of complex spatial analysis (especially those with high requirement on computational resources) while avoiding putting too much work load on the clients, especially the home computers of public participants.

On the client side, a standalone application is deployed on each client machine to replicate the collaboration tools and the 3D visualization module. A multi-cast communication is provided between the server and the clients to ensure consistency among multiple participants. More specifically, the client application contains a 3D viewer, a group of simple analysis functions, a set of collaboration tools, and a communication module.

The 3D viewer is implemented using a virtual globe plugin called Skyline Globe Viewer (http://www.skylinesoft.com). We choose Skyline Globe Viewer since it allows users to integrate their own remote sensing images (which may have a higher spatial resolution or are more timely than the default images provided) with the base map. However, other virtual globe plugins, such as Google Earth, could also be used if there are no specific requirements for the base images. The 3D viewer retrieves data from the data server and renders them as 3D models, labels and even diagrams.

The spatial analysis functions provide basic GIS tools, such as buffer and distance measurement, to assist the spatial decision-making process. When combined with more advanced geospatial services through the service module on the server side, the design system can provide strong spatial analysis capabilities while maintaining the flexibility to fit in different projects. The collaboration tools, which contain instant messages and sketches, are employed to facilitate the online collaborations.

The communication module is responsible for the correspondence between the client and the server. It sends the participant's messages and operation parameters to the server's synchronization application, receives the server's information, and passes such information to the 3D viewer and other components.

The three major functions of the design system are implemented based on the interactions of the client and server components discussed above. The leader-and-follower model has been used as the concurrency control mechanism for these three functions. The workflow (Figure 4) is designed as follows: (1) the leader makes an action (e.g., changing the view, drawing a sketch or performing an analysis) in the virtual environment; (2) the parameters of this action (e.g., view angles, the sketch's vertices, or the analysis parameters) are captured by the 3D viewer which then sends these parameters to the client's communication module; (3) the communication module transmits these parameters to the synchronization application on the register server; (4) the synchronization application then broadcasts these parameters to the communication module of all the followers; (5) the followers' communication module passes these parameters to the 3D viewer; (6) the 3D viewer replicates the same action.

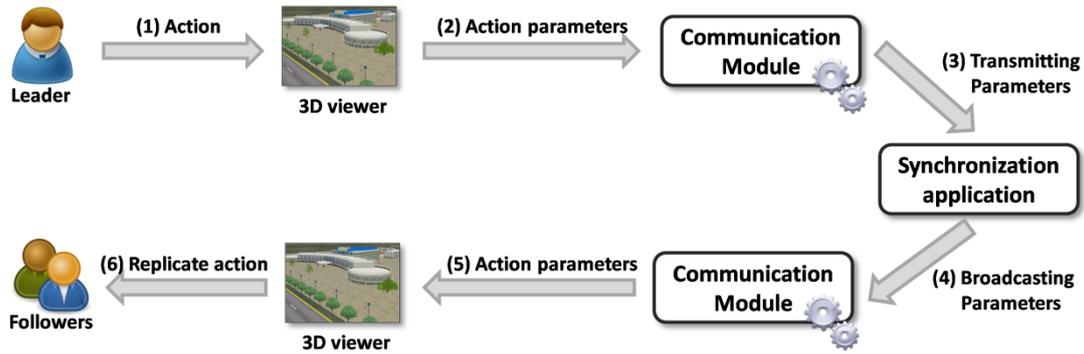

Figure 4: A general workflow of the synchronous collaborative functions

*4.2 The Review System*

The major purpose of the review system is to open the solution evaluation process to a broader public participants. To achieve this goal, the review system supports asynchronous collaboration which allows the interested public to login to the system at any time and any place to review the alternative solutions and make comments. Experts and participants who have joined the previous online meetings could also login to the review system to interact with other participants and answer questions.

*4.2.1 Functionalities*

An essential function in the review system is a geo-referenced discussion derived from the *Argumentation map* idea (Rinner 2001, Keßler *et al.* 2005, Rinner *et al.* 2008, Rinner and Bird 2009). Instead of linking a forum to a 2D map, the review system displays people's discussions as hyperlinks in a 3D environment to provide a seamless integration. When exploring in the virtual world, public participants can click on the hyperlinks to see other people's comments, and can also attach their own comments to geographic features.

*4.2.2 Architecture of the Review System*

The review system is designed as a browser-based application. Compared with the design system, the review system has a relatively simple architecture (Figure 5).

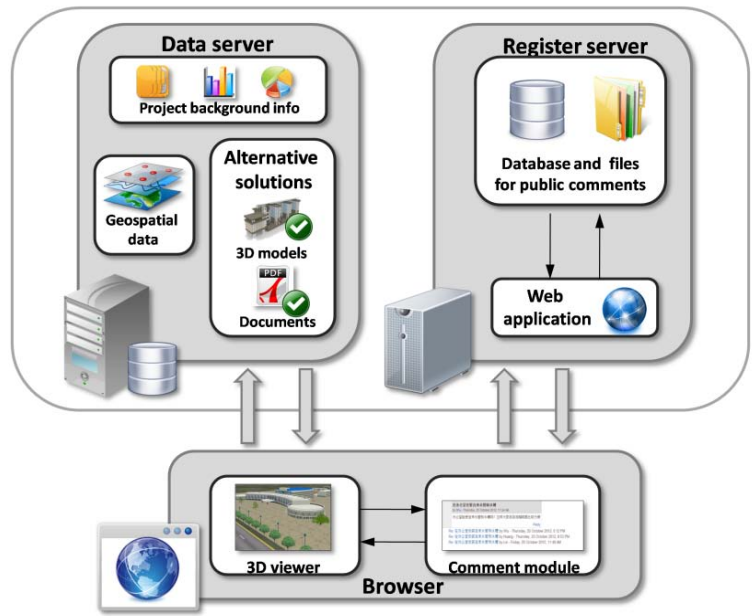

Figure 5: Architecture of the review system

As shown in Figure 5, the server side contains a data server and a register server. The data server publishes project data, geospatial data, and most importantly, the data of the alternative solutions. The register server hosts the database and files, which contain the information of public's comments. A web application is also running on the register server, and it bridges the client and the data of public comments.

On the client side, a 3D viewer and a comment module are embedded in a Web browser. Similar to the design system, Skyline Globe Viewer has been employed to deliver virtual globe-based visualization. Remote sensing images, 3D models, alternative solutions, public comments, and all other data are integrated in the virtual environment. The comment module allows public participants to input their opinions as well as to add geo-references.

*4.3 Net Participation Topology*

The presented framework supports collaboration among multiple participants holding different roles. Meanwhile, a variety of tools have been designed and embedded into the framework as communication channels. To give a whole picture of the participant roles, tools, and their relations, we draw a net participation topology (Figure 6).

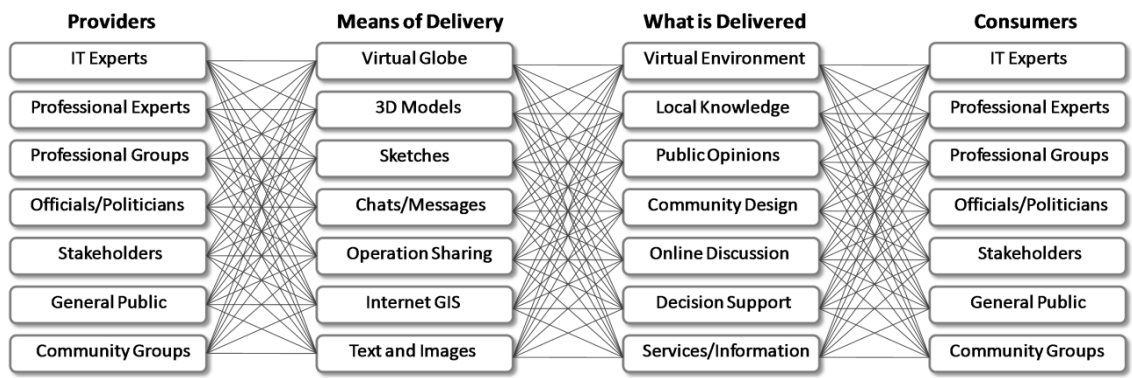

Figure 6: Net participation-- information providers, consumers, communication channels and delivered contents. Schema was proposed and used in (Hudson-Smith *et al.* 2002, Hanzl 2007)

The net participation topology has been proposed and used in previous studies (Hudson-Smith *et al.* 2002, Hanzl 2007). Learning from previous works, we list different types of participants at the two ends of the participation topology. In an interactive collaboration process, participants can be both information providers and consumers. For example, by providing background information about the project, a professional expert becomes the information provider; but he/she may also become an information consumer as stakeholders present information about the local community. Between the two ends of the topology, there are modules, tools and medias designed in the proposed framework, as well as the contents delivered by them. To give an example, the virtual globe, 3D models, text and images can deliver a virtual environment which contains different types of information and analysis services, while sketches, messages, and operation sharing can be used to express public opinions, local knowledge, and community design.

**5. Case Study**

The proposed system has been designed as a general framework that can be applied to public-involved and collaborative projects in different domains. For example, when integrated with land-use data and land-change models, it could support land-use planning and urban growth analysis. In this paper, we apply the system to an experiment of scene modeling.

Scene modeling is one of the areas to which collaborative platforms are often applied. For example, J. Dragonas and N. Doulamis (2009) designed a Web-based collaborative system which enables asynchronous collaboration among remotely distributed model designers. Bardis G. employed Multicriteria Decision Support and machine learning methods to personalize the collaboration process by identifying scene designers' preferences (Bardis 2009). While applied to a similar case, the system proposed in this study puts more emphasis on lowering the entry of public participation by using virtual globe-based 3D visualization. Besides, the system's collaborative functions are designed to primarily elicit participants' opinions instead of asking people to actually design 3D models.

*5.1 Experiment Overview*

This experiment convenes 21 graduate students from the Department of Geography at East China Normal University (ECNU), Shanghai, China to discuss a prospective plan for a new department building. While an experiment involving student volunteers may not be as objective as a real-world application, we hope it can, at least, demonstrate the system's functionalities and also provide a preliminary evaluation.

The experiment simulates the first four stages in Figure 1, i.e., problem definition, problem analysis, alternative solution generation, and solution evaluation. The 21 graduate students play the role of public participants. They are divided into two groups: the first group has 11 students who use the design system to collaborate in the first three stages and perform a first-step evaluation; the other 10 students use the review system to join in the evaluation stage to review and comment on the plans. The students in the first group can also login to the review system to join the discussions. Two authors of this research play as experts to coordinate this experiment. A group of 3D models, including buildings, trees, and facilities (e.g., street lamps), have been predesigned, and can be directly imported into the collaboration sessions.

### 5.2 Experiment procedure

The proposed framework has been implemented using Java as the programming language and Skyline Globe Viewer as the virtual globe platform. Two servers, a register server and a data server, have been used to deploy the system. Both of the two servers are composed of eight 2.0GHz-CPUs and 8GB memory, and running JDK 1.7 on Windows Server 2008. The design system's synchronization application and service module, as well as the review system's web application have been deployed on the register server.  The data server publishes the data used in the experiment, which include aerial remote sensing images (with a resolution of 0.5 meter) covering the campus area, 3D models created using 3ds Max, Shapefile vector layers, photo images, and text documents. Microsoft SQL Server 2008 has been used to store the comments from participants.

A tutorial session was held to help participants get familiarized with both of the two subsystems. The entire experiment was conducted in a distributed setting in which participants joined in the online collaboration sessions from different locations. In the problem definition stage, the experts introduced the objective and background of this experimental project. Vector data about the campus layout were imported, and text descriptions were displayed in popup windows (Figure 7). In the problem analysis stage, student participants and experts were engaged in a number of discussions, mostly focusing on students' expectations for a new department building. In the alternative solution generation stage, students and experts collaborated to design the building. Predesigned 3D models were employed in this stage (Figure 8a), and students were also allowed to upload their own models. Several simple scene modeling tools, such as the street tree tool (Figure 8b) and road pavement tool), have also been used. In the preliminary evaluation stage, the experts led student participants walk around in the virtual environment to assess the new department building. This assessment process was not limited to the building's external appearance but also extended to the inside (Figure 9). After the preliminary evaluation stage, the plan of the department building was published into the review system, and students in the second group could login into the system to review the building plan and give their comments (Figure 10). The comments were collected in a database, and another online session was held using the design system to address the raised issues. Necessary changes were made and explanations were given for the suggestions that were difficult to realize. The revised building plan was once again published into the review system, and most people were satisfied after this iteration. A meeting was held to learn participants' use experience, and all participants were asked to fill out a questionnaire to assess the usability of the system.

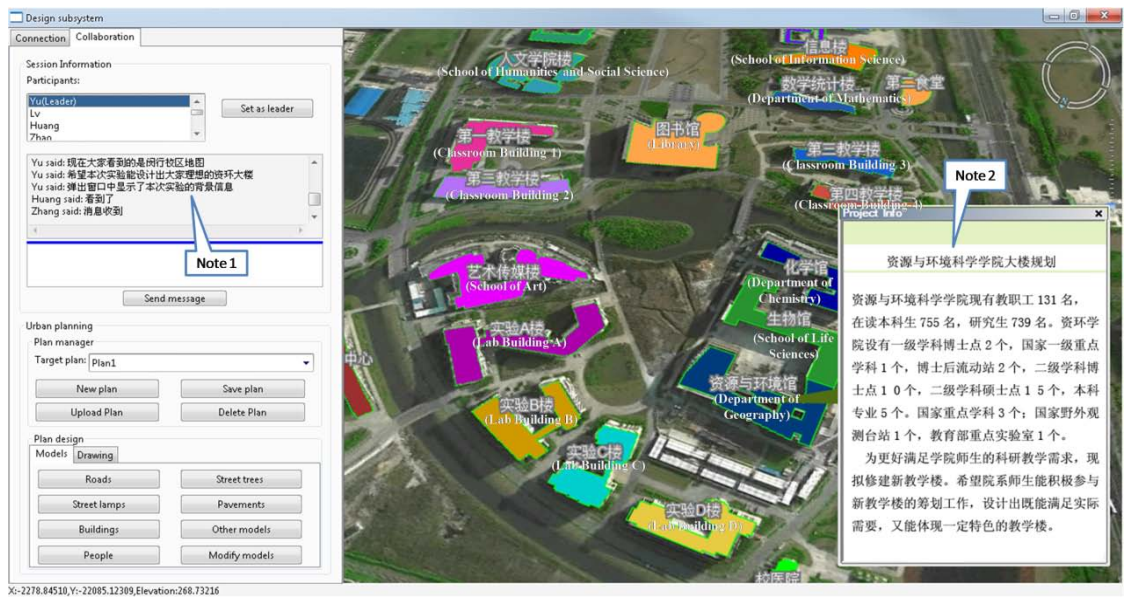

Figure 7: Popup windows and imported vector data in the virtual environment. English explanation of Note 1: participant Yu (role of expert) is introducing the project's background information, and participants Huang and Zhang reply to Yu's messages. English explanation of Note 2: the texts in the popup window give a general description about the ECNU Geography department (including the number of faculties, staffs, and students), and the purpose of this experiment.

### 5.3 Results and Discussions

The questionnaire, filled out by the participants, presented ten questions regarding the usability of the proposed framework. For each question, participants were asked how strongly they agree or disagree with the statement in a scale from 1 to 10. Participants can also choose "N/A" if they consider themselves as not qualified to answer this question (e.g. the participant did not use a particular tool). They can also add texts to explain their choices. The result of the questionnaire survey is summarized in Table 1 ( "N/A"s were not included to calculate the weighted averages).

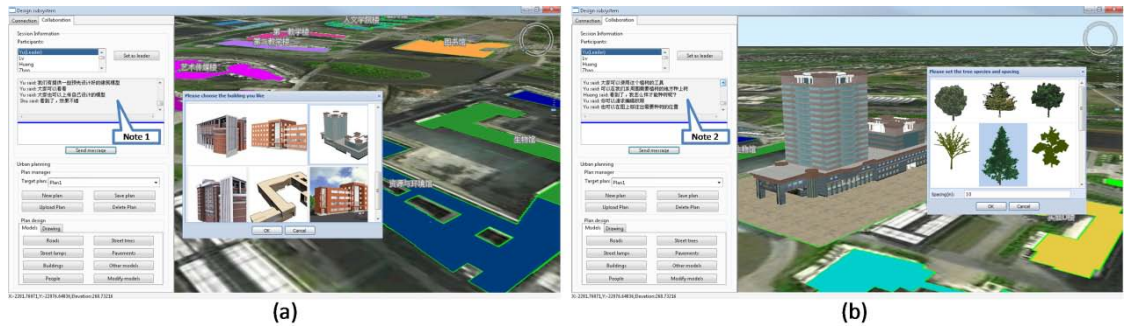

Figure 8: Alternative design choices provided for the participants: (a) alternative building styles; (b) alternative tree species. English explanation of Note 1: the expert participant is sending message to inform others that they can import the predesigned building models into the virtual environment. English explanation of Note 2: participant Huang is asking how to use the tool to plant trees in the virtual scene, and participant Yu is explaining to her.

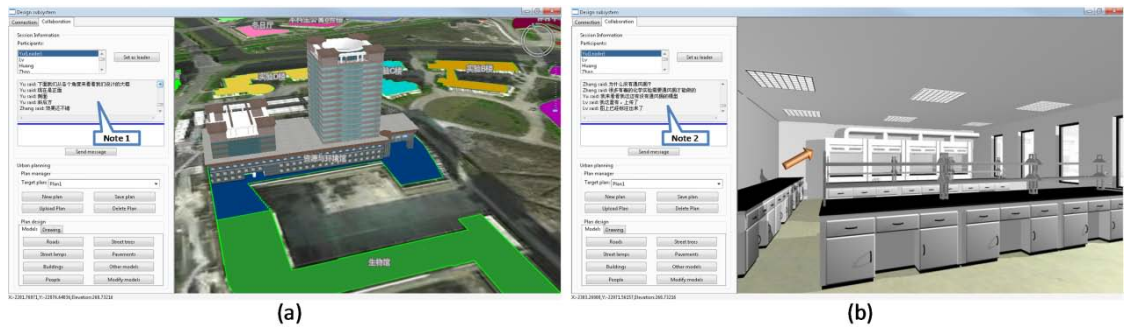

Figure 9: A first-step evaluation of the building plan using the design system; (a) an outside perspective; (b) an indoor perspective. English explanation of Note 1: one participant is leading others to walk around the newly designed building, and gives descriptions on different aspects of the building. English explanation of Note 2: participant Zhang is asking why stink cupboards are missing in the physical geography lab, and participant Lv is uploading a corresponding 3D model to fix this issue.

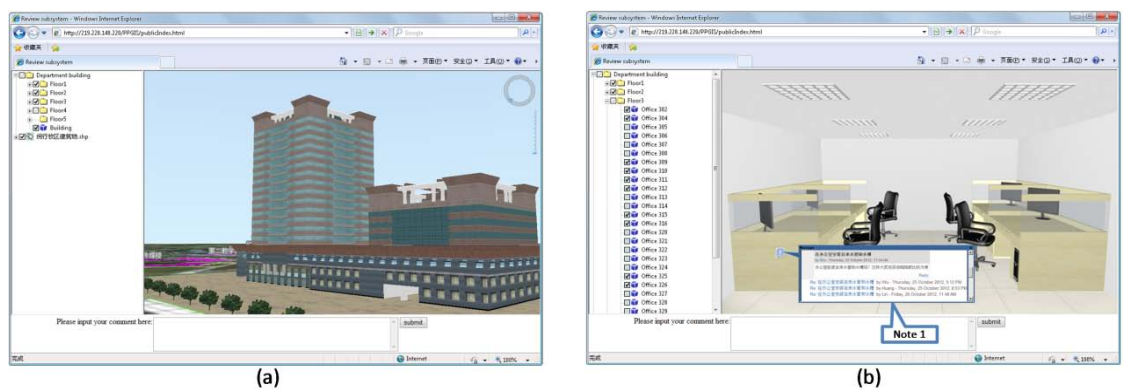

Figure 10: Evaluating the building plan in the review system; (a) an outdoor perspective; (b) an indoor perspective. English explanation of Note 1: participant Shu posts a comment in a office's virtual environment asking if it is possible to install water pipes and a sink in the office so that it would be more convenient to make coffee and tea.

Table 1. A summary of the questions and survey results (weighted averages and the number of "N/A"s)

| Question | Weighted average | Num of N/A |
|---|---|---|
| 1. With the assistance of the system, I feel I was strongly involved in the department building planning process. | 8.14 | 0 |
| 2. I can easily join the sessions at any location, and can review the plans and submit comments at any time and any place. | 9.33 | 0 |
| 3. This system effectively helped coordinate the multi-stage planning process. | 8.05 | 0 |
| 4. Compared with 2D architectural sketches, the virtual environment and 3D models provided by this system made it easier for me to understand the plan. | 8.33 | 3 |
| 5. Compared with 2D architectural sketches, the virtual environment and 3D models helped me understand the plan of the department building more comprehensively and in more details. | 6.94 | 4 |
| 6. This system encouraged me to explore the building in a virtual environment. | 9.43 | 0 |
| 7. Having the view synchronized helped me understand other participants' intentions and facilitated the collaboration. | 8.87 | 6 |
| 8. I can easily communicate with other participants using the chatting tool provided by the system. | 9.14 | 7 |
| 9. The drawing and geo-referencing tools helped me clearly specify the particular object that I was talking about. | 7.86 | 7 |
| 10. It was easy for me to use the editing tools to modify the models in the virtual scene. | 6.33 | 9 |

Question 1 to 3 regard the general performance of the collaborative 3D GIS in supporting a distributed multi-stage decision-making process. The high average scores indicate that most participants agreed that the system can facilitate public participation and collaboration in different stages.

Question 4 and 5 are designed to evaluate the effects of using a 3D visualization module in the system. A number of "N/A"s were observed, which could be explained by the comment from one participant who said he/she had no experience in 2D architectural sketches. Despite the number of "N/A"s, the high score in question 4 reflects a strong agreement in the capability of 3D visualization in facilitating understanding. However, the lower score in question 5 suggests that 3D models were not considered as very effective in achieving a comprehensive understanding. This seemingly paradox result could be explained by the fact noticed by the two authors who have joined in the experiment. Most 3D models used in the experiment were not accurate reflections of the real objects. Some parts of the models (e.g., the windows) were simply duplicated using the same image. These simplified models could give participants an impression that 3D visualization did not give an accurate representation of the reality.

Question 6 addresses the system's capability to attract public participants. The high score indicates that the collaborative 3D GIS encourages participants to explore in the virtual environment. The comment from one participant says "It was fun to walk around and even into the building, and see if it is like what you expected."

Question 7 to 10 are specifically designed to evaluate the effectiveness of the collaborative tools in the design system, which are primarily used by participants in the first group. A large number of "N/A"s are observed even though participants in the second group had also used these tools during the tutorial session. The high scores in question 7 and 8 indicate that the 3D view synchronization and the instant messaging tools have been considered as especially helpful in online collaboration.

The fair score in question 9 suggests that participants agree that the geo-referencing tools can help clarify the spatial meaning of text messages, but it may not be as helpful as the 3D view synchronization and chatting tools. This result is understandable as the experimental environment is the university campus with which most students are familiar. Consequently, there was less demand for a geo-reference tool to clarify the spatial meaning.

The score of question 10 is lower, which indicates that some of those tools still need improvements. During the tutorial session, some student participants experienced difficulty in using the model editing tools (e.g., some operations start with a sequence of single clicks and finish with a right click, which confuses participants). This low score suggests these difficulties also existed in the experiment.

**6. Conclusions and Future Work**

In this research, a multi-stage collaborative 3D GIS has been designed and implemented to support public participation. Based on a detailed literature review, we propose a dual-subsystem architecture consisting of a design system and a review system. The design system supports the stages of problem definition, problem analysis, alternative solution generation, and a first-step solution evaluation. It enables synchronous online meetings in which both decision makers and public participants can collaborate, design alternative solutions, and perform evaluations. The review system opens the solution evaluation process to a wide public participants by allowing interested citizens to review alternative solutions at any time and any location. Applying the concept of Digital Earth, the proposed framework employs a virtual globe as its major user interface. Such an interface integrates different types of project information into a unified virtual environment which is delivered to both experts and public participants through the Internet. While the proposed framework has been designed for general applications, we demonstrate its functionalities by a scene modeling case and receive generally positive feedbacks.

This research, however, still has some limitations that need to be addressed in future. One problem is the reconnection issue. During the experiment, one participant was disconnected from the session due to some network malfunctions. Although he was able to reconnect to the session after a while, the messages sent out by other participants when he was offline were lost. This issue could be solved by temporally storing participants' messages and operations on the register server, and send such message to users when they reconnect to the session. Secondly, while a volunteer-based experiment can provide some preliminary evaluation towards the system's usability, its result may not be as objective as real-world applications. Therefore, we also plan to find a suitable real project to further test the system.

**Acknowledgement**

The authors would like to thank the three anonymous reviewers for their constructive suggestions and comments. This work is supported by the National Natural Science Foundation of China (Grant No. 41001270), the Specialized Research Fund for the

Doctoral Program of Higher Education (Grant No. 20100076120017), the Specialized Research Fund of Key Lab of Geographic Information Science, Ministry of Education (Grant No. KLGIS2011C04), and the Grant from Shanghai Key Lab for Urban Ecological Processes and Eco-Restoration (Grant No. SHUES2012A03).